\documentstyle[prl,aps,twocolumn,floats]{revtex}
\begin{document}
\title{Fast quantum search algorithm and Bounds on it}
\author{Arun Kumar Pati}
\address{Theoretical Physics Division, 5th Floor, Central Complex,}
\address{Bhabha Atomic Research Centre, Mumbai - 400 085, INDIA.}
\date{\today}
\maketitle

\begin{abstract}
We recast Grover's  generalised  search  algorithm in a geometric
language even when the states are not  approximately  orthogonal.
We  provide  a  possible  search  algorithm based on an arbitrary
unitary  transformation  which  can  speed  up  the  steps  still
further.  We discuss the lower and upper bounds on the transition
matrix elements when the  unitary  operator  changes  with  time,
thereby  implying that quantum search process can not be too fast
or too slow. Quantum uncertainty relation puts bounds  on  search
process unlike classical cases.

\end{abstract}
PACS           NO:           03.65.Bz, 03.67.Lx.\\
email:apati@apsara.barc.ernet.in\\

\vskip 1cm
\par
After the seminal paper by Deutsch \cite{dh} the classical way of
envisaging computers has been changed. The information can now be
stored  not  only  in  two  distinct  bits  but  also in a linear
superposition of two bits, called a qubit. Any two-state  quantum
system  (e.g.  spin-1/2  particle,  spin  of  hydrogen  nuclei or
molecule)  can  in  principle  be  regarded  as  a qubit. Quantum
computer performs a given task by creating and  manipulating  the
superposition  of  various  computational  paths  resulting  in a
parallelism. For carrying out a specific task one has to  provide
a  quantum  algorithm  which is nothing but a sequence of unitary
transformations performed on the initial qubit. The final  output
is  read  by  projecting  on  to a particular state, i.e., when a
measurement is done in accordance  with  the  usual  concepts  of
quantum  theory.  That  quantum computers are much more efficient
than  any  other  classical  computers has been realised recently
after Shor's \cite{sr} algorithm  for  factoring  large  numbers.
Surprisingly,  Grover \cite{gr} has shown that a quantum computer
can search an unmarked item in a unsorted list of $N$ entries in
a step $O(\sqrt N)$, whereas a classical computer takes number of
steps $O(N)$. Bennett et al \cite{bt} have shown that search  can
not  be acomplished in less than $O(\sqrt N)$ steps. The original
proof used two elementary operations, one is Walsh-Hadamard (W-H)
transformation and  other  is  the  selective  inversion  of  the
amplitude  of  some  basis  state  (the  basis states belong to a
Hilbert space of dimension $2^n = N$) in a  quantum  computer.  A
Hamiltonian  description  of  search  algorithm has been given by
Farhi and Gutmann \cite{fg} showing that the time taken to  reach
a  target  state  is  $\sqrt  N/E$,  where  $E$  being the energy
eigenvalue. Grover's algorithm has been implemented by Chuang  et
al  \cite{ch}  using  nuclear  magnetic  resonance  where quantum
information was stored with a solution  of  chloroform  molecules
and  the  computation  has  been  carrried  out. It is found that
theory and experiment are in  good  agreement  with  each  other.
Zalka  \cite{za}  has  discussed  how  to obtain tight bounds on
search algorithm and limits on parallelizing Grover's  algorithm.
More  recently,  Grover  \cite{lgr}  has shown that a square root
reduction in computationla steps can be achieved by an  algorithm
which  uses  almost  any  unitary  transformation  instead of W-H
transformation and selective inversion  of  phase  of  the  qubit
basis states.

    In this letter we  recast  Grover's  recent  algorithm  in  a
geometric language, throwing some new light on the computational
process.  The  same  number of steps are obtained even when we do
not assume the  approximate  orthogonality  between  the  initial
state  and  the  adjoint  time  evolved target state. Further, we
provide a new search algorithm based only  on  a  single  unitary
operator  (in  contrast to composite unitary operator in Grover's
case) which  can  possibly  reduce  the  number  of  steps  still
further.  Finally,  we discuss both lower and upper bounds on the
transition matrix element which  governs  the  number  of  steps
involved  in  a  search  process,  when the unitary operator that
relates the initial and target state changes with time.  This  is
an important question in any search algorithm whose answer is not
known.  It  turns  out  that a quantum algorithm can not search a
particular entry too fast or too slow because there are upper and
lower bounds on the transition probabilities  related  to  energy
uncertainty in the qubits.

     The  quantum  search  algorithm  proceeds as follows. We are
given a function $f(x): X\rightarrow \{0,1\}$ defined on a domain
X  and  $x\in  X  = \{1,2,.....N\}$. This function has a non-zero
value  equal  to $1$ for some element $x = \tau$ and zero for all
other elements in the set $X$. The job is to find out $\tau$ when
we  have  no  information about the function $f(x)$. To implement
this  in  a  quantum  computer  we  require  a  quantum  register
(collection  of n qubits) whose state vector belongs to a Hilbert
space of dimension $2^n =  N$.  Then  each  basis  in  $N$  state
quantum  system  is  mapped  on  to  each  entry  in the set $X$.
Following Grover's prescription the system is initialised in such
a way that each basis state has same amplitude. Let  the  initial
state be $|\psi_i\rangle = |\gamma\rangle$. If the final (or target) state is
denoted as $|\psi_f\rangle = |\tau\rangle$ then any unitary operator can take
the  initial  state to the final state and if we do an experiment
to find the system in the final qubit state, the answer  will  be
with a probability $|\langle\psi_i|U|\psi_f\rangle|^2 = |\langle\tau|U|\gamma\rangle|^2 =
|U_{if}|^2$.    This    simple,    standard
probability  rule  of  quantum  theory  tells  us that one has to
repeat the experiment $1/|U_{if}|^2$ times to get successfully the
state $|\psi_f\rangle$. Grover's algorithm \cite{lgr} shows that  there
is a unitary operator $Q$ which can take $O(1/|U_{if}|)$ steps to
reach  the  final  state starting from an initial state. Here, we
recast  his  result  in  geometric   language   using   essential
structures from projective Hilbert space of the quantum computer.
In  quantum theory the state of system is not just a vector but a
ray. A ray is set of vectors which  differ  from  each  other  by
phase factors of unit modulus. Since any two-state is a qubit, we
say  set  of  qubits differing from each other by phase factors a
{\it raybit}. The projection of the {\it raybits} taken from  the
Hilbert  space  of the quantum register via a projection map give
us a projective Hilbert space.  If  the  original  Hilbert  space
$\cal  H$  has  dimension  $2^n = N$  then the projective Hilbert space
$\cal  P$  has  dimension  $N  -1$.  The  state  of  a  qubit  is
represented by a point on the projective Hilbert space (which can
be  a  $2(N  -1)$ dimensional unit sphere). This admits a natural
measure of distance called  Fubini-Study  distance  \cite{aa,ap}.
The  distance  between  any  two qubits $|\psi_1\rangle$ and $|\psi_2\rangle$
whose   projections   on   $\cal   P$   are   $\Pi(\psi_1)$   and
$\Pi(\psi_2)$, respectively can be defined as

\begin{equation}
d^2(\psi_1,\psi_2) = 4(1 - |\langle\psi_1|\psi_2\rangle|^2).
\end{equation}
This is gauge invariant and also invariant under all
unitary  and anti-unitary transformation defined over the Hilbert
space of the quantum computer. In search problem, first, we  want
to  reach  a state $|\psi_f'\rangle = U^{-1}|\psi_f\rangle$ from a initial state
$|\psi_i\rangle$. This means we  have  to  cover  the  Fubini-Study  distance
between  these  states  $U^{-1}|\psi_f\rangle$  and $|\psi_i\rangle$ which is
given  by  $d^2(\psi_i,\psi_f') = 4(1 - |U_{if}|^2)$. Application of
Grover's $Q$ operator creats  the  linear  superposition  of  the
state  state  $|\psi_i\rangle$  and  $|\psi_f'\rangle$.  It  is given by $Q =
-I_iU^{-1}I_fU$, where $I_i = (1- 2|\psi_i\rangle\langle\psi_i|)$ and $I_f  =
(1  -  2|\psi_f\rangle\langle\psi_f|)$ are unitary operators that changes the
sign of the basis states $|\psi_i\rangle$ and $|\psi_f\rangle$. In  one  step
the  initial  state changes to $|\psi_i^{(1)}\rangle = Q|\psi_i\rangle = (1 -
4|U_{if}|^2)|\psi_i\rangle  +  2   U_{if}|\psi_f'\rangle$.   Therefore,   the
distance  between  the  resulting  state and the initial state is
$d^2(\psi_i,\psi_i^{(1)}) = 16|U_{if}|^2$  (we  have  not  assumed
that  the  states  $|\psi_i\rangle$ and $|\psi_f'\rangle$ are necessarily
approximately orthogonal). This shows that in one application  of
$Q$ we can move the initial state a distance $4 |U_{if}|$. So the
serach  problem  tantamounts to  asking the question: in how many
applications of $Q$ we can  cover  a  distance  $d(\psi_i,\psi_f')$.
Clearly the number of steps $s$ involved will be given by

\begin{equation}
s  =  {d(\psi_i,\psi_f')  \over d(\psi_i, \psi_i^{(1)})
}  =  {1   \over   2} \sqrt \bigg({1\over   |U_{if}|^2}   -1 \bigg).
\end{equation}
Since  $|\psi_f\rangle$ can be obtained by single application of $U$ on
$|\psi_f'\rangle$, this shows that  we  can  reach  the  target  state  in
$O(1/|U_{if}|)$  steps  starting from a initial state $|\psi_i\rangle$,
which is Grover's algorithm.

    One  can  ask  is  there any other quantum algorithm that can
cover  the  Fubini-Study distance with much lesser steps? Indeed,
it is possible to speed up Grover's  algorithm  by  a  controlled
unitary  operation. The idea is to look for a unitary operator which
will move the initial vector by a larger amount so  that  we  can
reach  the  target  state  with lesser steps. Our algorithm is as
follows. Consider
a unitary  operator $V$ acting on the two-dimensional subspace spanned
by vectors $|\psi_i\rangle$ and $|\psi_f'\rangle$.

\begin{equation}
V = exp[-i{\alpha \over 2} (|\psi_f'\rangle\langle\psi_i| + |\psi_i\rangle\langle\psi|)]
\end{equation}
where  $\alpha  =  2  \sin^{-1}(|U_{if}|^p)$  and $p$ is a number
assumed to lye between $0 < p \leq 1$.  This  operator  preserves
the  two-dimensional vector space and when it acts on the initial
qubit state $|\psi_i\rangle$ creats linear  superposition  of  these
two states

\begin{equation}
|\psi_i^{(1)}\rangle = V|\psi_i\rangle = \cos(\sin^{-1}(|U_{if}|^p))|\psi_i\rangle + |U_{if}|^p |\psi_f'\rangle.
\end{equation}
This  is a controlled unitary operator depending on the parameter
$p$. Unlike Grover's unitary operator $Q$ (which is  a  composite
unitary  operator  constructed from three unitary operators) this
is  a  single  one.  Then  the  distance  between  these  vectors
$d(\psi_i,\psi_i^{(1)})  =  2|U_{if}|^p$ which we have covered in
one operation of $V$. Therefore, the total number  of  operations
required to cover a distance $d(\psi_i,\psi_f')$ is given by

\begin{equation}
s  =   {d(\psi_i,\psi)  \over d(\psi_i, \psi_i^{(1)})
}  =   {\sqrt (1 - |U_{if}|^2) \over |U_{if}|^p}
\end{equation}
which  is nothing but $O({1 \over |U_{if}|^p})$. If we choose the
initial angle $\theta$ such that $p = .5$, then  we  can  achieve
another  square  root  reduction  in number of steps in a quantum
search algorithm compared to Grover's algorithm.

   The above  result  can  also be obtained by dividing the angle
between the initial and target state and the angle by  which  the
initial  state rotates in one application of $V$. The appropriate
measure of angle between two non-orthogonal rays is the  Bargmann
\cite{bn}  angle.  If  $|\psi_1\rangle$ and $|\psi_2\rangle$ are two raybits,
then the Bargmann angle $\theta$ between them is defined as

\begin{equation}
|\langle\psi_1|\psi_2\rangle| = \cos{\theta \over 2},
\end{equation}
where  $\theta  \in  [0,\pi]$.  When  two   qubits   belongs   to
equivalence  classes  then  the  angle  is  zero,  therefore they
represent a single point on the projective Hilbert space  of  the
quantum  register.  When two qubits are orthogonal they represent
diametrically opposite points on the $2(N-1)$ dimensional sphere.
Quantum mechanical search problem is then equivalent to searching
an appropriate point on sphere. The Bargmann angle between initial raybit
$\Pi(\psi_i)$  and  target  raybit  $\Pi(\psi_f')$ is $\theta = 2
\cos^{-1}|U_{if}|$.  The  search  algorithm tells us how much one
has to rotate the point $\Pi(\psi_i)$ so as to reach $\Pi(\psi_f')$
in a fewer steps. By one application of  the  controlled  unitary
operator we rotate $\Pi(\psi_i)$ by an angle
$\theta_0 = 2\cos^{-1}(|\langle\psi_i|\psi_i^{(1)}\rangle|) = 2\sin^{-1}|U_{if}|^p$
Therefore,  the  number  of times one needs to apply the operator
$V$ in order to reach the point $\Pi(\psi_f')$ is given by

\begin{equation}
s = {\theta \over \theta_0} = {\cos^{-1}|U_{if}| \over \sin^{-1}|U_{if}|^p},
\end{equation}
which is $O({1 \over  \sin^{-1}|U_{if}|^p})$.  When  the  quantity
$|U_{if}|$ is much less than unity, this reduces to result
$O({1  \over  |U_{if}|^p})$.  If  we  are dealing with exahustive
search starting with an initial state $|\psi_i\rangle =  |0\rangle$  and  the
unitary  transformation  $U$  is  the  W-H  transformation (for a
single qubit W-H transformation is $U = U_{WH} = {1  \over  \sqrt
2} \left( \begin{array}{ccc}
1 & 1\\
1 & -1\\ \end{array} \right)$  with  $U_{WH}^2 = 1$), then $U_{if}$ is ${1 \over \sqrt
N}$ for any  final  target  state.  Grover's  algorithm  requires
$O(\sqrt  N)$  steps,  whereas  our  algorithm takes $O(N^{p/2})$
steps. Advantage of the new algorithm is its ability to implement
it  using  netron spins or photon polarisations schemes. All that
is needed is sequence of desired rotations that can creat  linear
superposition of the states $|\psi_i\rangle$ and $|\psi_f'\rangle$. The index
$p$ of the algorithm controlls the amount by  which  the  initial
qubit has to be rotated.

An important question in any fast quantum search algorithm is how
``fast''  is  the  ``fast''?  Is  it  true  that  with   improved
algorithms  the  number  of  steps  can  be  drastically reduced,
thereby throwing away all classical algorithms.  Is  there  any
lower  limit to the number of steps involved in search algorithm.
Similarly, one can ask if a given algorithm  is  not  optimal  is
there an upper limit to the number of steps? The answers to these
question will be of fundamental importance in  future.  Here,  we
show   that   quantum   mechanical   uncertainty   relation  puts
restriction on the number of steps required in a search  process.
In  search process we remove the assumption that has been made in
Grover's as well as in ours and that  is  the  operator  $U$  and
$U^{-1}$  do  not remain constant during search. When the unitary
operator changes with time we can think that quantum  computation
is  governed  by a Hamiltonian $H(t) = i\hbar \dot{U}U^{-1}$. The
initial qubit state $|\psi_i\rangle$ changes with  time  and  the  time
evolution  is  prescribed  by  $|\psi_t\rangle  = U(t)|\psi_i\rangle$. Let us
consider the transition probability matrix element $|U_{if}(t)|^2$
and its time evolution:

\begin{eqnarray}
&&{d \over dt}|U_{if}(t)|^2   = \langle\psi_f|U(t)\rho_iU^{-1}(t)|\psi_f\rangle \nonumber\\
&& = {1 \over \hbar} |\langle\psi_f|[H(t), \rho(t)]|\psi_f\rangle|
\end{eqnarray}
where  $\rho_i  =  |\psi_i\rangle\langle\psi_i|$  and  $\rho(t) = U(t) \rho_i
U^{-1}(t)$.  Then  using  the  generalised  uncertainty  relation
between two non-commuting operators we have

\begin{equation}
|{d \over dt}|U_{if}(t)|^2| \leq
{2     \over     \hbar}     \Delta     H    ~~(|U_{if}(t)|^2    -
|U_{if}(t)|^4)^{1/2},
\end{equation}
where $\Delta H$ is the energy uncertainty in  the  target  state
$|\psi_f\rangle$.  Applying the properties of differential inequalities
\cite{pf} we have $P_{-}(t) \leq P(t) \leq P_+(t)$ (for  $t  \geq
0$) where $P(t) = |U_{if}(t)|^2$ and
$P_{-}(t)$  and $P_+(t)$ are the maximal and minimal solutions of
the initial value problem during computation.

\begin{equation}
|{d \over dt}P_{\pm}(t)| = \pm 2 {\Delta H \over \hbar} (P_{\pm}(t) - P_{\pm}^2(t))^{1/2}
\end{equation}
The above equation can be cast in terms of Bargmann angle as $ {d
\over dt}\theta_{\pm}(t)
=   \pm  2{\Delta H \over  \hbar}$, where $\theta_{\pm}(t)$ is defined through
$P_{\pm}(t) = \cos^2 {\theta_{\pm}(t) \over 2}$.
This  results  in  $P_+(t)  =
\cos^2({\theta(0) \over 2} +  \int {\Delta H \over \hbar} dt)$ and
$P_-(t) = \cos^2({\theta(0) \over  2}  -  \int  {\Delta H \over
\hbar})$, where $\theta(0) = 2 \cos^{-1}(\langle\psi_f|\psi_i\rangle|)$ is the
angle between the initial and the target qubit state.  Therefore,
the transition matrix element that governs the number of steps in
a quantum search algorithm is given by

\begin{eqnarray}
&&|U_{if}(t)| \leq \cos({\theta(0) \over 2} +  \int {\Delta H \over \hbar} dt) \nonumber\\
&&|U_{if}(t)| \geq \cos({\theta(0) \over  2}  -  \int  {\Delta H \over
\hbar}dt )
\end{eqnarray}
In  Grover's  algorithm  the  number  of  steps is still given by
$O(1/|U_{if}(t)|)$ when unitary operator changes  with  time.  In
fast  search  algorithm such as ours the number of steps is again
given by $O(1/|U_{if}(t)|^p)$. Note that we are  not  considering
the change in $U$ due to small perturbations in the operation but
purely  dynamical changes due to some Hamiltonian that drives the
qubit states. Therefore, the same derivation will go  through  in
Grover's case even in time-dependent situation. Although, we have
not proved that for perturbative changes the same form will hold,
we  can  remark  some  interesting  point  here. If we change the
Hamiltonian  by some amount, then $|U_{if}(t)|)$ remains the same
provided the added Hamiltonian commutes with  the  original  one.
Also,  under  symmetry  transformation  of  the  Hamiltonian that
causes initial qubits  to  change,  the  quantity  $|U_{if}(t)|)$
remains  invariant  due  to Wigner theorem. So, Grover's and ours
algorithm are stable for such situations. For small  perturbative
changes  of  the  unitary  operator preliminary calculation shows
that the number of steps may  change  and  this  result  will  be
reported in future.

These  number  of  steps  $s$  in Grover's algorithm has both a
lower and an upper bound.

\begin{eqnarray}
&& s \geq \sec({\theta(0) \over 2} +  \int {\Delta H \over \hbar} dt) \nonumber\\
&& s \leq \sec({\theta(0) \over  2}  -  \int  {\Delta H \over
\hbar} )
\end{eqnarray}

In the search algorithm provided here, the same is true. The bounds are given by

\begin{eqnarray}
&& s \geq \sec^p({\theta(0) \over 2} +  \int {\Delta H \over \hbar} dt) \nonumber\\
&& s \leq \sec^p({\theta(0) \over  2}  - \int  {\Delta H \over
\hbar} )
\end{eqnarray}

This  clearly  shows  that although quantum computation takes the
advantage  of interference and speed up the search algorithm, the
quantum uncertainty in the energy of the target state puts bounds
on the number of steps involved. As a result  of  this  a  search
algorithm based on quantum
principle can not be too fast or too slow-- a unique feature of the uncertain
quantum world.

Before   concluding the paper, we remark on a problem that may be
encountered in time-dependent search algorithms  and  it  is  the
``slippage  of  state''  problem.  When  the unitary operator $U$
changes with time so also the operator $Q$ and $V$. Now if  there
is  a  time  $T  =  st$,  where $t$ is time required to apply the
operator once, such that the state after $s$ number of steps i.e.
the state $|\psi_i^{(s)}\rangle = Q^s|\psi_i\rangle$ becomes  the
initial  state  upto  an overall phase, then the quantum computer
would not search further. It has to restart again a  fresh  (i.e.
reset) from the initial state, because the state
 $|\psi_i\rangle$   and  $|\psi_i^{(s)}\rangle$  belong  to  same
raybit. If this happens, then the search algorithm is not  useful
and  hence  one has to change the operator $Q$ in such a way that
at time $T = st$ the slipage does not occur.

    To   conclude   this   paper,   we   provide   a  geometrical
understanding  of  the  Grover's  search   algorithm   based   on
projective  Hilbert  space of qubits. By calculating Fubini-Study
distance we  arrived  at  the  earlier  result  of  Grover  which
provides  a  geometrical  understanding  to  search  process.  We
provided a fast  search  algorithm  based  on  a  single  unitary
operator  which improves the number of steps still further. It is
a controlled algorithm,  which  we  believe  can  be  implemented
although  we have not explicitly constructed an example. When the
unitary  operator  changes  with  time we have shown that quantum
uncertainty  relation  puts  lower  and  upper  bound  on  search
process. Therefore a quantum search process cannot be
too  fast  or too slow--an important point in quantum computation. But
certainly  it  has  advantage over classical computers and future
will decide its fate.\\

\vskip 1cm

{\bf   Acknowledgement:}   I   thank   H.  D.  Parab  for  useful
discussions. I wish to thank L. K. Grover for  useful  electronic
discussions and going through my paper carefully.

\renewcommand{\baselinestretch}{1}

\end{document}